# The Future Time Traveller Project: Career Guidance on Future Skills, Jobs and Career Prospects of Generation Z through a Game-Based Virtual World Environment

Michalis Xenos[1,2], Catherine Christodoulopoulou[1], Andreas Mallas[1,2], and John Garofalakis[1,2]

[1] CTI- Directorate of Telematics and Applications for Regional Development, "D. Maritsas" Building, N. Kazantzaki str., GR26504 Rion, GREECE
[2] Computer Engineering and Informatics Dep., University of Patras, Rio Campus, Building B, GR26500, GREECE
`{xenos, mallas, garofala}@ceid.upatras.gr, christod@westgate.gr`

**Abstract.** Future Time Traveller is a European project that aims at transforming career guidance of generation Z through an innovative, games-based scenario approach and to prepare the next generation for the jobs of the future. The project objective is to foster innovative thinking and future-oriented mindset of young people, through an innovative game-based virtual world environment. This environment helps them explore the future world, understand the trends that shape the future world of work, the emerging jobs, and the skills they will require. The Future Time Traveller project is implemented by a team of experts in 7 European countries (Bulgaria, Germany, Greece, Italy, Poland, Portugal, and United Kingdom). The project target groups include young people (generation Z), career guidance practitioners and experts, and policymakers. This paper presents, in brief, the Future Time Traveller project and introduces the reader to the main features and functionalities of the 3-dimensional virtual world and the games developed in this environment.

## 1 The Future Time Traveller Project

The Future Time Traveller Project [1] (called Future hereinafter) is a project running from December 2017 that will be completed in November 2020 (duration 36 months). The project is coordinated by Business Foundation for Education (Bulgaria) and the team includes Computer Technology Institute and Press "Diophantus" (Greece), Aspire-Igen group (UK), European Board for Certified Counselors (Portugal), Centro Italiano per l' Apprendimento Permanente (Italy), Institute of Learning Innovation (Germany), and University of Lodz (Poland).

The project outcomes include a) a policy mapping study focusing on young people's awareness about future jobs, career guidance practitioners' capacity for innovation, and the role of policy support for sustainable, quality, innovative career guidance development, b) the creation of a game-based 3-dimentional virtual world environment for exploring the future trends, jobs and skills, and c) the creation of an e-book with the best practices of innovative career services, as well as the creation of a policy evidence report about the project outcomes. This paper focuses on the presentation of the 3-dimensional virtual world environment, which is discussed in the following sections.

## 2      The 3-Dimensional Virtual World Environment

The concept of gamification is relatively new and there are theoretical frameworks of unified discourses [2], while in the field of education the use of games has a rapid growth [3]. Examples of similar works using 3-dimensional (3D) virtual worlds and games for educational and informative purposes are the use of such worlds for simulated events related to software project management [4] and for informing children for the risks of fires and floods using a safe simulated environment [5].

The 3D virtual world environment of the Future project is based on Open Simulator [6], which is a free, open-source, 3D application server that allows the creation of 3D virtual worlds, where multiple users can simultaneously be present. These virtual worlds can be accessed through various open-source clients and can remain private, behind the firewalls, or become public. Open Simulator is written in C# and its framework is designed to be easily extensible through external modules.

Using this concept, the participants use a client program to connect to the Future 3D virtual world. The client program enables users to interact with each other through avatars (human-shaped computer figures). A single user account may have only one avatar at a time, although the appearance of this avatar can change between as many different forms as the user wishes. Avatar forms, like almost everything else, can be created by the user with the use of a high-level interface. The avatar can be dressed up as the user desires, interact with other objects and avatars and even animate some physical movements of a user's choice. The avatars interact with each other, as well as with non-player characters (NPGs) and with objects within the 3D virtual world, allowing the creation of complex games.

## 3      The Game Concept, Skills, and Competences

When the players enter the 3D virtual world they are located in the information centre (see Fig.1) where they are able to choose their language of preference (the game is multilingual supporting all partner languages) and to familiarise themselves with the environment (i.e. how to navigate, how to interact, how to change views). After they feel comfortable with the environment, they can start playing.

The game starts in the year 2020. In the local library, the player understands about the arrival of a time machine and receives its mission to act as an ambassador for the future. Upon correctly sorting various sources of information about future jobs, the player teleports to the year 2050 (see Fig 2). There, the player participates in several missions designed to test their knowledge about the future of work, explores future jobs and defines the necessary skills and competences they require. Furthermore, they investigate the future of jobs (i.e. which jobs will remain, which jobs will change fundamentally or disappear in the future). Before returning to the present, the player consolidates the received knowledge in a message to humanity.

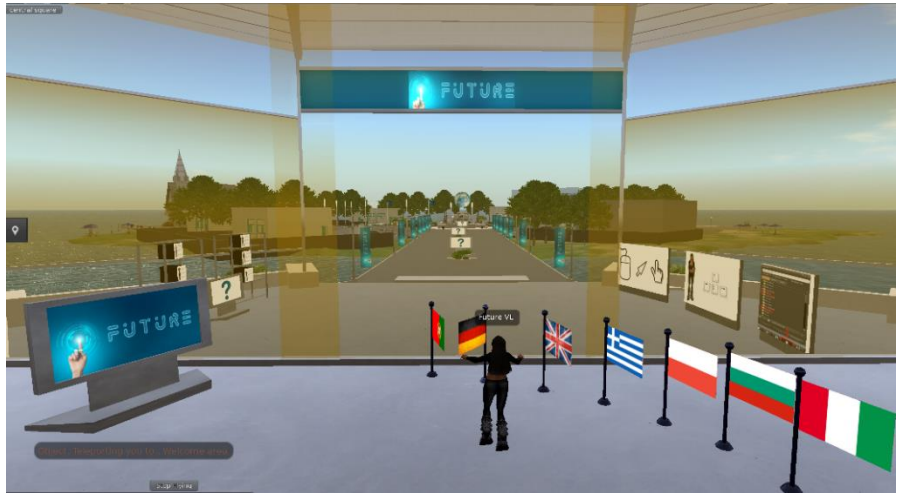

**Fig. 1.** The entry point into the 3D virtual world of the Future project, language selection and basic information.

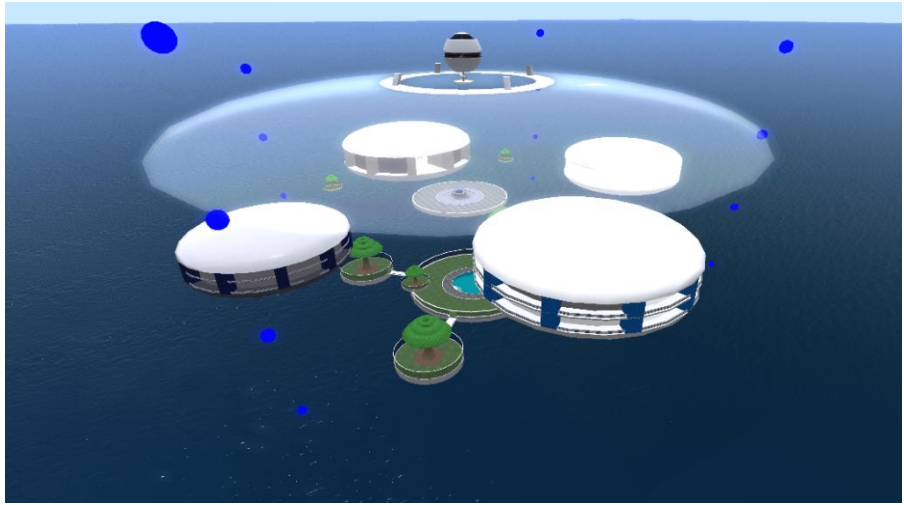

**Fig. 2.** Overview of the future world of 2050.

Back to 2020, (see Fig 3) the player must create a short, original description of a future job that does not exist yet. In the final self-reflection task, the player will synthe-size the main ideas and insides provoked by the game, in a personal "Message to myself in the future", which will be delivered in a chosen moment in time.

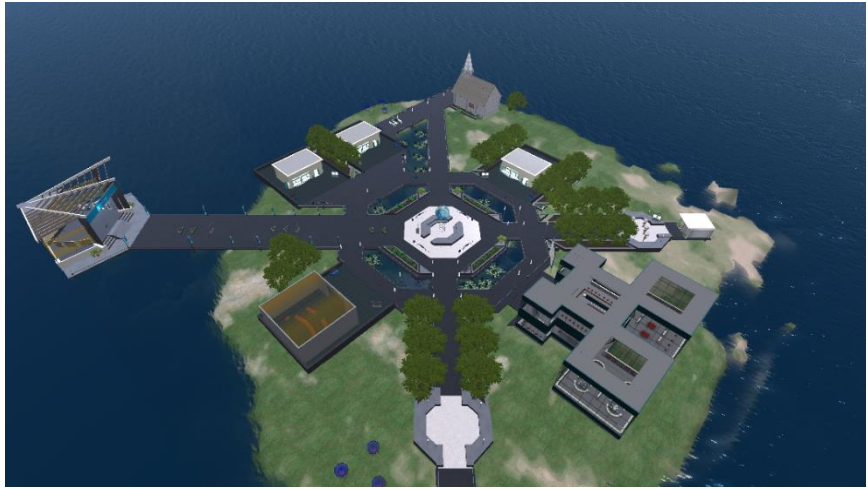

**Fig. 3.** Overview of the world of 2020 with all the game areas.

The game scenarios in the Future project 3D virtual world use elements of escape rooms, web quests, treasure hunting, strategy, and adventure games. Throughout their learning journey the users will have to complete various missions, which will help them:

- Improve knowledge of labour market trends that shape the future world of work.
- Raise awareness of the challenges of technological development and their impact on emerging jobs.
- Improve the understanding of the required skills of the future.
- Gain insight into different career opportunities.
- Understand their own role for mastering the future.
- Develop career management competencies, critical thinking, decision making, problem-solving, creativity and flexibility, self-reflectivity, etc.

## 4  Conclusion and Future Work

The 3D virtual world of the Future project is already in the piloting phase in all participating countries (Bulgaria, Germany, Greece, Italy, Poland, Portugal, and United Kingdom) that will allow users from all target groups to navigate through the environment and to participate in the various in-world games. During the pilot phase, users from all countries will be able to co-exist in the virtual world, to play the games, discuss their experiences and to be informed about new developments and outcomes of the Future project.

Future work includes the further piloting of the 3D virtual world environment, informing the various project stakeholders (e.g. young people known as generation Z, career guidance practitioners and experts, policymakers) in all partner counties and in various European countries and developing an e-book with the best practices of innovative career services, as well as the creation of a policy evidence report about the project outcomes. Updates of the project outcomes will be available at the Future project homepage [1].

## Acknowledgement

The work presented in this paper is co-funded by the Erasmus+ Programme of the European Union, project no: 590221-EPP-1-2017-1-BG-EPPKA3-PI-FORWARD. The authors would like to thank everyone in this project from all the participant organisations, as well as the numerous users of the project outcomes.